\documentclass[prl,twocolumn,showpacs,amssymb]{revtex4}
\usepackage{graphicx}% Include figure files
\usepackage{dcolumn}% Align table columns on decimal point
\usepackage{bm}% bold math
\newcommand{\be}{{\bf e}}
\newcommand{\br}{{\bf r}}
\newcommand{\bB}{{\bf B}}
\newcommand{\bA}{{\bf A}}
\newcommand{\bH}{{\bf H}}
\newcommand{\bk}{{\bf k}}
\newcommand{\bq}{{\bf q}}

\newcommand{\bQ}{{\bf Q}}

\begin{document}

\title{Unscreened universality class for superconductors with columnar
disorder}

\author{Anders Vestergren$^{(a)}$, Jack Lidmar$^{(b)}$, and Mats
Wallin$^{(a)}$}

\affiliation{(a) Condensed Matter Theory, Royal Institute of
Technology, SCFAB, SE-106 91 Stockholm, Sweden\\ (b) Department of
Physics, Stockholm University, SCFAB, SE-106 91 Stockholm, Sweden}

%\date{\today}
\date{\today, {\em Physical Review B in press}}
\begin{abstract}
The phase transition in a model for vortex lines in high temperature
superconductors with columnar defects, i.e., linearly correlated
quenched random disorder, is studied with finite size scaling and
Monte Carlo simulations.  Previous studies of critical properties have
mainly focused on the limit of strongly screened vortex line
interactions.  Here the opposite limit of weak screening is
considered.  The simulation results provide evidence for a new
universality class, with new critical exponents that differ from the
case of strong screening of the vortex interaction.  In particular,
scaling is anisotropic and characterized by a nontrivial value of the
anisotropy exponent $\zeta=\nu_\parallel/\nu_\perp$.  The exponents we
find, $\zeta = 1.25 \pm 0.1, \nu_\perp=1.0 \pm 0.1, z = 1.95 \pm 0.1$,
are similar to certain experimental results for YBCO.
\end{abstract}

\maketitle

Columnar defects have proven very effective at pinning vortex lines in
high temperature superconductors (SC), resulting in significant
increases in the critical magnetic fields and currents~\cite{Civale}.
The columnar defects are often produced as damage tracks by
irradiating the sample with high-energy heavy ions.  Such correlated
disorder strongly affects the phase transition from the SC to normal
state and there is an ongoing theoretical and experimental
effort to investigate the phase diagram and properties of the
transition
\cite{Nelson,Wallin-Girvin,Lidmar-Wallin,Lee,Blatter,Zlatko,Phuoc,Phuoc2,Klein,Soret,Smith,Smith2,Jiang,Reed,grigera,Kwok}.
Below a critical temperature the vortices freeze into a glassy SC
state, known as a Bose glass, which replaces the Abrikosov vortex
lattice of the clean system.  Upon heating the system undergoes a
continuous phase transition to a resistive vortex liquid state.  The
transition is characterized by critical exponents, that have been
measured in a number of experiments.  The critical exponents are
universal, i.e., they are common for all systems in the same
universality class, and depend only on general features like the range
of the interaction and distribution of the disorder.  In spite of the
fact that the screening length $\lambda$ in high temperature SC is
quite large, most theoretical work on columnar defects has
concentrated on strongly screened systems with short range
interactions\cite{dirtybosons,Wallin-Girvin,Lidmar-Wallin}.  In
general, long range interactions often change the universality class of
a transition, compared to short range interactions.  It is therefore
natural to study the universality class with long range interactions.
In this paper we present a systematic study of the SC transition in
systems with columnar disorder, an applied magnetic field, and weak
screening, and compare the results with experiments.

The superconducting phase transition with columnar disorder is also
interesting from the perspective of quantum phase transitions in
systems with disorder
\cite{Nelson,Wallin-Girvin,dirtybosons,Fisher-Grinstein-Girvin,Herbut}.
In fact, statistical mechanics of vortex lines in three-dimensions is
closely analogous to world-lines of quantum bosons in (2+1)
dimensions, where the vortex phase transition corresponds to the
zero-temperature boson localization by substrate disorder.  The
quantum bosons have static disorder in imaginary time, corresponding
to columnar disorder for the vortex lines.  The quantum dynamical
exponent $\zeta$ (relating the diverging time and length scales at the
transition) translates into an anisotropic scaling behavior for the
vortex line problem, where correlations along the columnar defects
diverge with a different rate than the perpendicular ones when
approaching the Bose glass transition.

The vortex phase transition with columnar disorder has previously been
studied for the case of strongly screened vortex interactions
\cite{Wallin-Girvin,Lidmar-Wallin}.  There the correlation length
exponent $\nu$, anisotropy exponent $\zeta$, and dynamic exponent $z$,
take the values $\nu \approx 1, \zeta=2, z \approx 4.6$
\cite{Lidmar-Wallin}.  The result $\zeta = 2$ follows if the
compressibility is nonsingular through the transition, and is hence
believed to be exact.  The effect of long range interactions in
systems with correlated disorder has been considered in the context of
the superconductor-insulator quantum phase transitions of disordered
ultrathin films \cite{Fisher-Grinstein-Girvin,dirtybosons,Herbut}.
For long range planar $1/r$-interactions, that act only between world
line segments in the same $xy$-plane, and thus is completely local in
the $z$-direction, the system is incompressible, with $\nu \approx 1$
and $\zeta=1$ \cite{dirtybosons}.  Early simulations
\cite{Lee,Cha-Girvin} of more realistic models (from the vortex point
of view) show evidence for a non-trivial value of the anisotropy
exponent close to $1$.  Some experiments~\cite{Jiang,Reed} have been
interpreted as belonging to an incompressible universality class with
$\zeta \approx 1$.

The main task in this paper is to examine the superconducting phase
transition with columnar disorder in the limit where the screening
length is very long, such that any effects of screening of the vortex
interaction can be effectively neglected.  In other words we set
$\lambda \to \infty$, and thus examine a no-screening fixed point for
the superconducting phase transition.  The screening length in the HTS
is typically $\lambda \sim 1000$ {\AA} so for most of the phase
diagram the assumption of no screening is reasonable.  We stress that
one possibility is that the true asymptotic critical behavior is
effectively described by a fixed point with a screened interaction.
However, even if this is the case, we expect that the scaling
properties upon approaching the transition is controlled by the
unscreened model in a large region of the parameter space, before an
eventual crossover to the true critical regime.  In comparison, this
occurs in clean systems without an applied field \cite{FFH}.  Our
model also assumes no fluctuations in the amplitude of the SC order
parameter (London approximation), which is valid in a large region of
the phase diagram well below $H_{c2}$ \cite{Blatter}.  The interaction
is taken to be a fully isotropic, three-dimensional (3D) long range
interaction between vortex lines, that should apply for fairly
isotropic systems.  The form of the interaction makes our model {\em
different} from most of the previously studied Boson-like models, with
a planar interaction between the vortex lines.  We consider only the
case when both the magnetic field and all the columnar defects are
aligned in the $z$-direction.  The anisotropy due to the field and the
columnar disorder allows for the possibility of an anisotropic
correlation volume, where the correlation length in different
directions can diverge with different exponents.  In addition we
consider dynamical aspects of the problem, which do not have a
counterpart in the quantum boson problem, and compute the dynamical
exponent $z$.  This assumes that MC dynamics for vortex lines can be
equated with real dynamics, which should apply close to the transition
where the dynamics is slow and over-damped \cite{dynamics}.  Our
values for $\nu,\zeta,z$ differ from those of the previously studied
models, which suggests that the present model belongs to a new,
distinct universality class.

Our starting point is the Ginzburg-Landau theory for the
superconducting complex scalar order parameter field,
$\psi(\br)=|\psi| \exp i\theta(\br)$,
\begin{eqnarray}
H&=&\int d^d\br\, 
\left[ \left| \left( \nabla - \frac{2\pi i}{\Phi_0} 
\bA \right) \psi \right|^2
+ \alpha |\psi |^2 \right.
\nonumber
\\
&&
\left. + \frac{1}{2} \beta |\psi |^4
+ \frac{\bB^2}{8\pi} - \frac{\bB \cdot \bH}{4\pi} \right]
\end{eqnarray}
where $\Phi_0$ is the magnetic flux quantum, $\bB=\nabla \times \bA$
is the magnetic flux density and $\bA$ is the magnetic vector
potential, and $\bH$ is the applied magnetic field.  By fairly
standard manipulations \cite{helicity} this model can be transformed
into a model involving only the vortex degrees of freedom.  In the
London approximation, i.e., neglecting fluctuations in the amplitude
of $\psi(\br)$, a discretized form of the vortex Hamiltonian reads
\begin{equation}
H=\frac{1}{2} \sum_{\br, \br'} V(\br-\br') \bq(\br) \cdot
\bq(\br') + \frac{1}{2} \sum_\br D(\br_\perp) q_z(\br)^2,
\label{H}
\end{equation}
where the sums over $\br$ run through all sites on a simple cubic
lattice with $\Omega=L\times L\times L_z$ sites, and periodic boundary
conditions in all three directions.  The vortex line variables are
specified by an integer vector field $\bq(\br)$, such that the
$\mu=x,y,z$ component is the vorticity on the link from the site $\br$
to $\br+\be_\mu$.  The partition function is $Z={\rm Tr} \exp(-H/T)$,
where $T$ is the temperature, and ${\rm Tr}$ denotes the sum over all
possible integers $q_\mu$, subject to the constraint that the discrete
divergence $\nabla \cdot \bq=0$ on all sites, i.e., the vortex lines
have no free ends.  The vortex-vortex interaction is given by
\begin{equation}
V(\br)=\frac{K}{\Omega}\sum_{\bk} \frac{e^{i \bk \cdot
\br}}{\sum_\mu (2-2\cos k_\mu ) + \lambda^{-2}},
\end{equation}
where we choose units such that $K=1$, and consider the limit of no
screening, corresponding to $\lambda=\infty$.  Randomness is included
in the vortex model (\ref{H}) in the second term in the form of a
random core energy, which corresponds roughly to a random
$\alpha(\br)$ in the GL model.  In real systems, each columnar defect
gives approximately the same pinning energy, while they are located at
random positions.  On the lattice it is more practical to instead
insert one column on each link of the lattice in the $z$-direction,
but with a random pinning energy.  We use a uniform distribution of
random pinning energies in the interval $0 \le D(\br_\perp) \le 0.8$,
where each $D(\br_\perp)$ is constant in the $z$-direction to model
correlated disorder.  We model the net applied magnetic field as a
fixed number of $L^2/4$ vortex lines penetrating the system in the
$z$-direction, corresponding to quarter filling, i.e.\ one vortex on
every fourth link in the $z$-direction.  To test for universality of
our critical exponents, we also studied the case of $L^2/2$ vortex
lines penetrating the system in $z$ direction and disorder strength $0
\le D(\br_\perp) \le 1.0$.

The MC trial moves consist of attempts to insert vortex line loops of
random orientation on randomly selected plaquettes of the lattice.
One MC sweep consists of one attempt on average to insert a loop on
every plaquette, which we define as one MC time step, $\Delta t=1$.
The attempts are accepted with probability $1/(1+\exp \Delta E/T)$,
where $\Delta E$ is the energy change for inserting the loop.  The
initial vortex configuration is taken to be a regular lattice of
straight lines.  To approach equilibrium we discard about $2 \cdot
10^4$ MC sweeps ($6.5 \cdot 10^4$ for the resistivity) before any
measurements are taken, followed by equally many sweeps for collecting
data.  To verify that the warm-up time is long enough, we tried to
vary the warm-up time between 5000 and $10^5$ sweeps for a single
temperature, close to $T_c$, which gave no significant differences in
the final results.  The results were averaged over up to 1000-2000
samples of the disorder potential.  For the simulation of static
quantities we use an exchange MC algorithm to speed up convergence
\cite{exchange}.  Note that the exchange method can not be used for
dynamic quantities, since it involves large nonlocal moves in phase
space.  In the following we denote thermal averages by $\langle \dots
\rangle$ and disorder averages by $[ \dots ]$.

In the simulation, the helicity modulus and the RMS current is
obtained by the following procedure \cite{helicity}.  An extra term
$H_Q=\frac{K}{2 \Omega} \bQ^2$ is included in the Hamiltonian, where
$Q_\mu$ is the total projected area of vortex loops added during the
simulation \cite{helicity}.  The helicity modulus in the direction
$\mu$ ($\mu=x,z$) is then given by
\begin{equation}
\label{upsilon}
Y_\mu=1-\frac{K}{\Omega T} [ \langle Q_\mu^2 \rangle - \langle
Q_\mu^\alpha \rangle \langle Q_\mu^\beta \rangle ],
\end{equation}
and the RMS current density is given by $J_\mu=\frac{K}{\Omega} [
\langle Q_\mu^\alpha \rangle\langle Q_\mu^\beta \rangle ]^{1/2}$,
where we use two different replicas in our simulations, denoted
$\alpha$ and $\beta$, to avoid any bias in the expectation values.  We
also calculate the linear resistivity $\rho$, in which case $H_Q$ is
not included in $H$, by evaluating the Kubo formula \cite{Kubo}
\begin{equation}
\label{rho} 
R_\mu=\frac{1}{2T} \sum_{t=-t_0}^{t_0} [ \langle
V_\mu(t) V_\mu(0) \rangle ],
\end{equation}
where $t$ is MC time and $t_0 \to \infty$, and the voltage is
$V_\mu\sim\Delta Q_\mu$.  $\Delta Q_\mu$ is the net change in the
projected vortex loop area during a sweep.  In practice the summation
time $t_0$ is chosen large enough that the resistivity is independent
of $t_0$.

Next we consider the anisotropic finite size scaling relations used to
extract critical properties from the MC data
\cite{Wallin-Girvin,Lidmar-Wallin}.  At the transition temperature
$T_c$ the correlation length in the $xy$-planes, $\xi$, in the
$z$-direction, $\xi_z$, and the correlation time, $\tau$, are assumed
to diverge as $\xi \sim |T-T_c|^{-\nu}, \quad \xi_z \sim \xi^\zeta,
\quad \tau \sim \xi^z$.  The anisotropic finite size scaling ansatz
\cite{Lidmar-Wallin} for the helicity modulus in $z$ direction is
\begin{equation}
\label{fss_Uz}  Y_z=L^{1-d+\zeta} f_z(L^{1/\nu}(T-T_c),L_z/L^\zeta),
\end{equation}
where $d=3$ is the spatial dimensionality, and $f_z$ is a scaling
function (all scaling functions will from now on be suppressed).
Similarly, for the current density in the $x$-direction, we have
\begin{equation}
\label{fss_J}
J_x \sim L^{2-d-\zeta}.
\end{equation}
The linear resistivity is given by $\rho=E/J$, where $E$ is the
electric field and $J$ the current density, and scales as
\begin{equation}
\label{fss_rho}
\rho_x \sim L^{d-3+\zeta-z},\quad
\rho_z \sim L^{d-1-\zeta-z}.
\end{equation}

\begin{figure}[htb]
\resizebox{!}{7cm}{\includegraphics{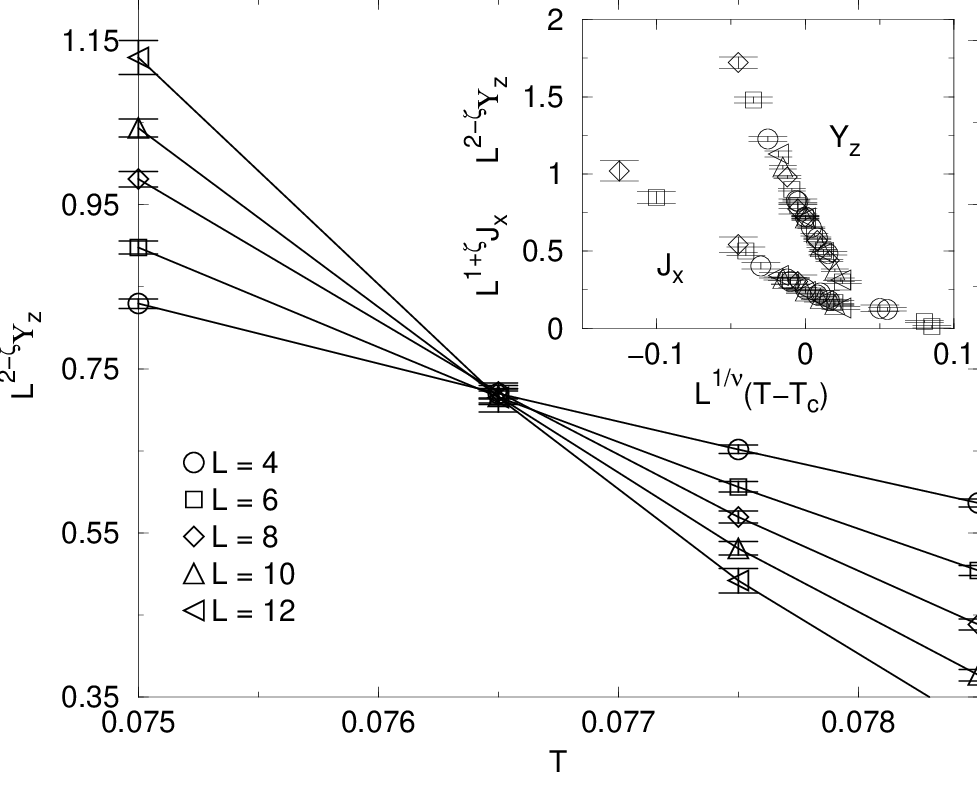}}
\caption{MC data for the helicity modulus in the $z$-direction vs.\
temperature for different system sizes $L$.  $T_c \approx 0.0765$ is
estimated as the temperature where all curves intersect.  Inset:
Finite size scaling collapse of the data for fillings $f=1/2$ and
$f=1/4$, and for two disorder strengths (see text) using $\nu = 1.0$,
$\zeta=1.25$.}
\label{helicity}
\end{figure}

\begin{figure}[htb]
\resizebox{!}{7cm}{\includegraphics{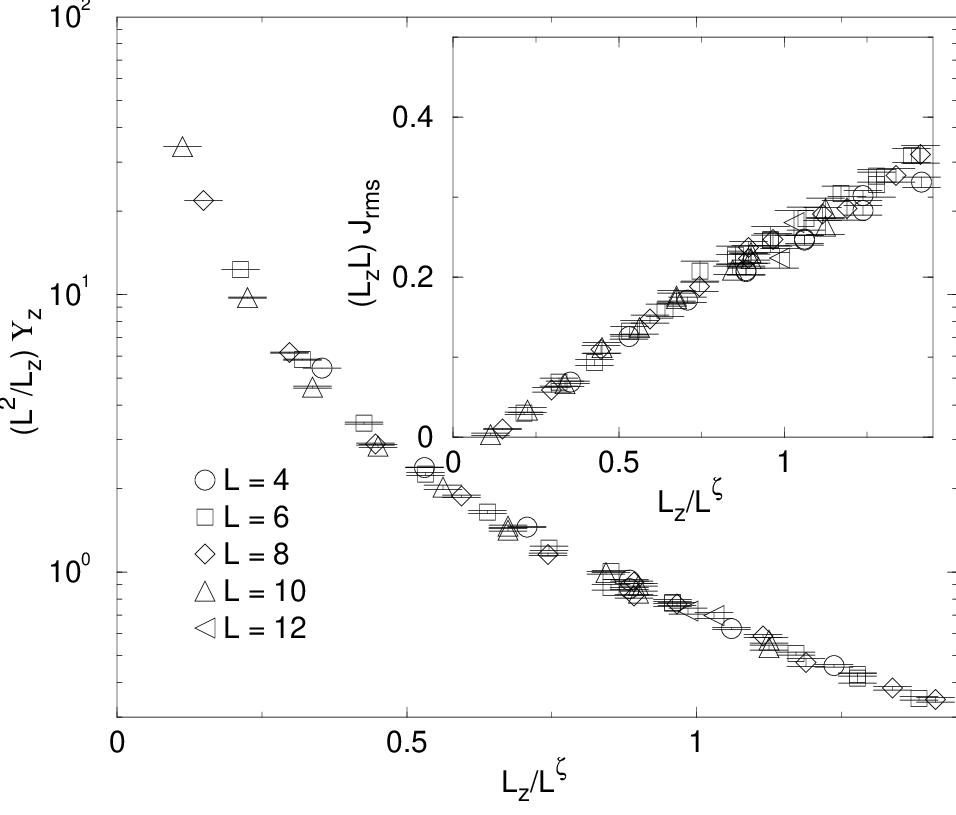}}
\caption{Finite size scaling collapse according to Eqs.\
(\ref{fss_Uz}) and (\ref{fss_J}) of data for the helicity modulus in
$z$-direction [main part] and RMS current in perpendicular directions
[inset] at $T=T_c$.  The best collapse is obtained for $T_c \approx
0.0765$, $\zeta \approx 1.25$. }
\label{collapse}
\end{figure}

In the simulations we consider a whole range of system sizes $L_z$ for
each $L$, since the anisotropy exponent $\zeta$ that enters the aspect
ratio, $L_z \sim L^\zeta$ is a priori unknown.  For non-integral
values of $L_z$, we simulate two systems with nearest integer $L_z$
and interpolate the results using linear interpolation.  For the
resistivity, both linear and logarithmic interpolations were tested
and agree within error bars.  Once we locate the correct value for
$\zeta$, the argument $L_z/L^\zeta$ in the scaling functions can be
made constant by setting $L_z \propto L^\zeta$.

We determine the critical temperature $T_c$ for the phase transition
and the critical exponents from MC data for the helicity modulus in
Eq.\ (\ref{upsilon}), by fits to the finite size scaling form in Eq.\
(\ref{fss_Uz}).  Figure \ref{helicity} shows our estimate of $T_c$
using the scaling form of the helicity modulus in the $z$-direction.
By selecting the value for $\zeta$ that gives the best common
intersection point, we obtain $\zeta=1.25 \pm 0.1$ and $T_c=0.0765$,
where the error bar on $\zeta$ is estimated by the interval outside
which scaling gets considerably worse.  The RMS current density in
$x$-direction, $J_x$, can also be used to calculate $T_c$.  The result
agrees, within the error bars, with the result from $Y_z$.

Right at $T_c$ it is possible to obtain a data collapse when plotting
various quantities as a function of $L_z/L^\zeta$.  Figure
\ref{collapse} show such plots for $(L^2/L_z) Y_z$ and $(L_z L)
J_\mathrm{rms}$.  Adjusting $T_c$ and $\zeta$ until the best collapse
is obtained gives identical estimates, within the error bars, as the
previous method.  This clearly demonstrates that the scaling is
anisotropic.

The inset of Fig.\ \ref{helicity} shows a determination of the
correlation length exponent $\nu$ using finite size scaling, for data
from two different fillings, $f=1/2$ and $f=1/4$, and two different
disorder strengths.  We use fits to Eq.\ (\ref{fss_Uz}) to obtain a
data collapse for different system sizes ($L=4-8$ for $f=1/2$,
$L=8-12$ for $f=1/4$) over an entire temperature interval around
$T_c$.  Both $T_c$ and $\nu$ were free parameters in the fit resulting
in $\nu = 1.0 \pm 0.1$, $T_c=0.0765$ for $f=1/4$, and $T_c=0.0747$ for
$f=1/2$.  The data collapse to a common universal scaling function
indicates that the results are indeed universal.

\begin{figure}[htb]
\resizebox{!}{7cm}{\includegraphics{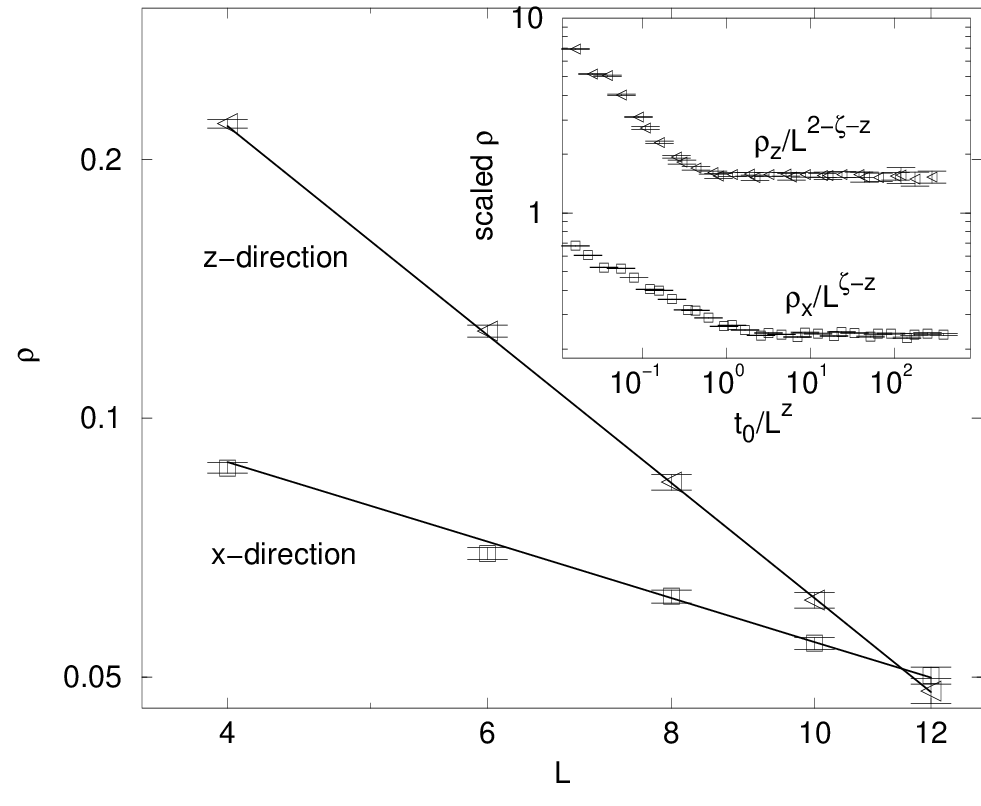}}
\caption{MC data for resistivity in the $x$ and $z$-directions for
different system sizes $L$ at $T=T_c$.  Straight lines are power law
fits to the data points, with exponents $z-\zeta = 0.65$ and
$z+\zeta-2 =1.28$.  Inset: Finite size scaling plot of the resistivity
as a function of the summation time $t_0$.  }
\label{rhofig}
\end{figure}

Next we study the critical MC dynamics of the model.  We calculate the
resistivity $\rho$ in $x$ and $z$-directions from the Kubo formula in
Eq.\ (\ref{rho}) at $T=T_c=0.0765$.  This gives us both a useful
consistency test of the value $\zeta \approx 1.25$ found above, and a
value for the dynamic exponent $z$.  To verify that the summation time
in Eq.\ (\ref{rho}) is long enough, we plot $\rho$ vs.\ $t_0$ in the
inset of Fig.\ \ref{rhofig}.  We also note that the correlation time
is at most $\sim 10^3$ sweeps, which is much less than the
equilibration times used in the simulations.  Figure \ref{rhofig}
shows the resistivity in the $x$ and $z$-directions as a function of
system size.  We calculate $\zeta$ and $z$ by making a power law fit
of the data points in the figure to Eq.\ (\ref{fss_rho}).  This gives
$\zeta = 1.3 \pm 0.1, z = 1.95 \pm 0.1$, where the error bars are
estimated by the bootstrap method \cite{numrec}, in good agreement
with $\zeta\approx 1.25$ found above.

\begin{table}
\caption{Summary of a few selected experiments and simulations of high
temperature superconductors with columnar disorder.}
\begin{tabular}{lccc} \hline \hline
Experiment/Simulation & $\nu$ & $\zeta$ & $z$ 
\\ 
\hline 
\\ 
${\rm (K,Ba)BiO_3}$ \cite{Klein} & 
$1.1 \pm 0.1$ & $\approx 2$ & $5.3 \pm 0.3$ 
\\ 
${\rm Tl_2Ba_2CaCu_2O_8}$ \cite{Phuoc} & $1.1 \pm 0.2$ & $1.9
\pm 0.2$ & $4.9 \pm 0.2$ 
\\ 
${\rm Bi_2Sr_2Ca_{1-x}Y_xCu_2O_8}$\cite{Soret} & $1.04 \pm 0.06$ & $\approx
2$ & $5.28 \pm 0.05$ 
\\ 
${\rm YBa_2Cu_3O_7}$ \cite{Jiang} & $\approx 1.0$ & $\approx 1.1$ & $\approx
2.2$ 
\\ 
${\rm YBa_2Cu_3O_7}$ \cite{Reed} & $0.9 \pm 0.2$ & $1.2\pm0.2$ 
& $2.3 \pm 0.3$ 
\\
Simulations ($\lambda \to 0$) \cite{dirtybosons,Lidmar-Wallin} &
$\approx 1.0 $ & 2 & $4.6 \pm 0.3$ 
\\ 
{\bf This work} ($\lambda \to \infty$) & $1.0 \pm 
0.1$ & $1.25 \pm 0.1$ & $1.95 \pm 0.1$ \\
\label{experiments}
\end{tabular}
\end{table}

Finally we will compare our findings with some other results.  The
results for critical exponents, $\zeta = 1.25 \pm 0.1, \nu=1.0 \pm
0.1, z = 1.95 \pm 0.1$, imply that the linear resistivity scales as
$\rho \sim |T-T_c|^s$ with $s_x=\nu(z+3-d-\zeta) \approx 0.7,
s_z=\nu(z+1-d+\zeta)\approx 1.3$.  For the nonlinear current-voltage
characteristic at $T=T_c$, we have $E \sim J^p$ with
$p_x=(1+z)/(d+\zeta-2)\approx 1.3, p_z=(\zeta+z)/(d-1) \approx 1.6$,
for $d=3$.  Table \ref{experiments} shows a comparison between
simulation results and results from a selection of transport
experiments where the critical exponents studied in this paper have
been measured.  In the table we observe that several experiments agree
very well with the previously known exponents for screened vortex
interactions.  Notably, however, the table also shows that the results
from two of the experiments on YBCO agree quite well with our
exponents for unscreened 3D interactions, which suggests that they may
effectively belong to the new universality class considered here.
Naively one may expect that the present results describe the
transition when the screening length is much longer than the vortex
spacing.  For e.g.\ YBCO, where $\lambda \sim 1400$ {\AA}, this
corresponds to $B \gtrsim 0.1$ T.  The precise location of the
crossover between weak and strong screening is unclear, as well as the
precise role of anisotropy and other parameters, and would be
interesting to investigate further both experimentally and
theoretically.

In summary, we obtained scaling properties and numerical results for
critical exponents that apply for the superconducting transition in
systems with columnar defects in the limit of a long screening length,
and compared with experiments.  The critical exponents for this
universality class differ considerably from the strongly screened case
\cite{Wallin-Girvin,Lidmar-Wallin}, and also from the case of planar
$1/r$-interactions \cite{dirtybosons}.  In particular, the new
anisotropy exponent differs from the value $\zeta=1$ assigned to the
universality class of incompressible dirty bosons with planar long
range interactions \cite{Fisher-Grinstein-Girvin,dirtybosons,Herbut}.
Further work is motivated in order to further clarify the origin of
the scaling properties obtained in different experiments, and when the
different models apply.  Experimental studies to look for a crossover
between unscreened and screened scaling behavior would also be
interesting.

We acknowledge illuminating discussions with George Crabtree and
Stephen Teitel.  This work was supported by the Swedish Research
Council, STINT, PDC, and the G{\"o}ran Gustafsson foundation.

\end{document}